\documentclass[copyright]{eptcs}
\providecommand{\volume}{118}
\providecommand{\anno}{2011}

\usepackage{inputenc}
\usepackage{url}
\usepackage{hyperref}
\usepackage{graphicx}
\usepackage{mathptmx}
\usepackage{helvet}
\usepackage{listings}
\usepackage{latexsym}
\usepackage{amsthm}
\usepackage{amssymb}
\usepackage{empheq}
\usepackage{lastpage}
\usepackage{fancyhdr}
\usepackage{footmisc}

\usepackage{rotating}
\usepackage{pdflscape}
\usepackage{algorithm2e}

\newtheorem{theorem}{Theorem}

\input{diagrams}
\diagramstyle[h=2mm,w=3mm,PostScript=dvips
]

\newcommand{\msg}[1]{\mathsf{msg}(#1)}
\newcommand{\dmsg}[1]{\mathsf{dmsg}(#1)}

\newcommand{\kind}[1]{\ensuremath{\mathsf{#1}}}

\newcommand{\pk}[1]{\operatorname{\kind{pubk}}(#1)}

\newcommand{\skel}{\ensuremath{\mathbb{A}}}

\newcommand{\bnd}{\mathcal{B}}
\newcommand{\bndC}{\mathcal{C}}

\newcommand{\cons}{\,{\hat{\ }}\,}

\newarrow{StrNext}      ===={=>}
\newarrow{DotStrNext}   ....{=>}
\newarrow{DotTo}   ....{>}
\newarrow{Bond} ----{->}
\newarrow{DashTo}{}{dash}{}{dash}>
\newarrow{MapTo}      |---{->}

\newarrow{StrNext}      ===={=>}
\newarrow{Bond} ----{->}
\newarrow{DashTo}{}{dash}{}{dash}>

\newcommand{\enc}[2]{\{\!\!|#1|\!\!\}_{#2}}

\newcommand{\term}{\kind{msg\/}}

\newcommand{\sembrack}[1]{[\![#1]\!]}

{\bfseries}{\itshape}
{\bfseries}{\itshape}
{\bfseries}{\upshape}
{\bfseries}{\itshape}

\newcommand{\lts}[1]{\ \stackrel{#1}\longrightarrow\ }
\newcommand{\msgbox}[2]{[#1]_{#2}}
\newcommand{\interact}[4]{#1\rightarrow#2:\mathsf{#3}\langle#4\rangle}

\newcommand{\pp}{\mathrel{\boldsymbol{\mathord{\mid}}}}

\newcommand{\pfx}{\mathbf.\ }
\newcommand{\INACT}{\mathbf0 }

\newcommand{\who}{\mathsf{who}}

\newcommand{\Rule}[2]{\displaystyle{\frac{#1}{#2}}}
\newcommand{\Did}[1]{(\textsc{#1})}
\newcommand{\NI}{\noindent}

\newcommand{\op}[1]{{\sf #1}}

\newcommand{\semproves}[3]{#1\,\models\,#2\,\rhd\,#3}

\newtheorem{definition}{Definition}

\def\titlerunning{Execution Models for Choreographies and Cryptoprotocols}
\def\authorrunning{Carbone and Guttman}

\title{Execution Models for Choreographies and Cryptoprotocols}

\author{Marco Carbone\thanks{The author was partially supported by EPSRC grant EP/F002114}\\
IT University of Copenhagen\\
Copenhagen, Denmark\\
\url{carbonem@itu.dk}\\
\and
Joshua Guttman{}\\
Worcester Polytechnic Institute\\
Worcester, MA, United States\\
\url{guttman@wpi.edu}\\
}

\begin{document}
\maketitle

%
\begin{abstract}
A choreography describes a transaction in which several principals
interact.  Since choreographies frequently describe business processes
affecting substantial assets, we need a security infrastructure in
order to implement them safely.  As part of a line of work devoted to
generating cryptoprotocols from choreographies, we focus here on the
execution models suited to the two levels.   

We give a strand-style semantics for choreographies, and propose a
special execution model in which choreography-level messages are
faithfully delivered exactly once.  We adapt this model to handle
multiparty protocols in which some participants may be compromised.  

At level of cryptoprotocols, we use the standard Dolev-Yao execution
model, with one alteration.  Since many implementations use a "nonce
cache" to discard multiply delivered messages, we provide a semantics
for at-most-once delivery.   
\end{abstract}

\section{Introduction}
\label{sec:introduction}

Choreographies are global descriptions of transactions including
business or financial transactions.  They describe the intertwined
behavior of several principals as they negotiate some agreement
and--frequently--commit some state change.  A key idea is
\emph{end-point projection}~\cite{carbone.honda.yoshida:esop07}, which
converts a global description into a set of descriptions that
determine the local behavior of the individual participants in a
choreography.  Conversely, \emph{global synthesis} of a choreography
from local behaviors is also sometimes possible, i.e.~meshing a set of
local behaviors into a comprehensive global
description~\cite{MostrousYoshidaHonda09}.

Because these transactions may transfer sums of money and other
objects of value, or may communicate sensitive information among the
principals, they require a security infrastructure.  It would be
desirable to synthesize a cryptographic protocol directly from a
choreography description, to control how adversaries can interfere
with transactions among compliant principals.  Corin et
al.~\cite{CDFBL08} have made a substantial start on this problem, with
further advances described in~\cite{BhargavanEtAl09}.  However, many
questions remain, for instance how to optimize the generated
cryptographic protocols, how best to establish that they are always
correct, and indeed how best to define their correctness.

This last question concerns how to state what control the protocol
should provide, against adversaries trying to interfere with
transactions.  It is a substantial question because the execution
model that choreographies use is quite distant from the execution
model cryptographic protocols are designed to cope with.  For
instance, choreographies use an execution model---or communication
model---in which messages are never received by any party other than
the intended recipient, or if the formalism represents channels, they
are received only over the channel.  Moreover, messages are always
delivered if the recipient is willing to receive the message.
Messages are delivered only if they were sent, and specifically only
if they were sent by the expected peer.  Finally, they are delivered
only once.  These aspects of the model mean that confidentiality and
integrity properties are built into the underlying assumptions.  A
security infrastructure is intended to justify exactly these
assumptions, i.e.~to provide a set of behaviors in which these
assumptions are satisfied.

Naturally, these behaviors must be achieved within an underlying model
in which the adversary is much stronger.  In this model---typically
called the \emph{Dolev-Yao model}, after a paper~\cite{DolevYao83} in
which Dolev and Yao formalized ideas suggested by Needham and
Schroeder~\cite{NeedhamSchroeder78}---all messages may be received by
the adversary, so that confidentiality needs to be achieved by
encryption.  They may be delivered zero times, once, or repeatedly,
and they may be misdelivered to the wrong participant.  When
delivered, a message may appear to come from a participant that did
not send it.  The adversary may alter messages in transit, including
applying cryptographic operations using keys that he knows, or may
obtain by manipulating the protocol.

Digital signatures may be used to notify a recipient reliably of the
source of a message (and of the integrity of its contents).  Symmetric
encryption may also be used to ensure authenticity:  a recipient knows
that the encrypted message was prepared by a party that knew the
secret key, and intended it for a peer that also knew the secret key.
Nonces, which are simply randomly chosen bitstrings, may be used to
ensure freshness.  The principal $P$ that chose a nonce knows, when
receiving a message containing it, that the nonce was inserted after
$P$ chose it.  Moreover, if $P$ engages in many sessions and
associates a different nonce with each, $P$ can ensure that messages
containing one nonce cannot be misdirected to a session using a
different nonce. 

In this paper, we begin the process of relating the Dolev-Yao model of
execution to the choreography execution model.  This is a key step in
generating cryptographic protocols and proving them faithful to the
intent of the choreography.  In particular, we represent the two
execution models using the strand space
model~\cite{strandspaces,GuttmanEtAl05}.

\paragraph{Goals of this Paper.} We provide a few definitions and
an example to indicate how the strand space framework can relate
choreographies to the cryptographic protocols that implement them.

In particular, we consider a very simple choreography language, and
provide a semantics for it as a set of ``abstract bundles.''  That is,
each session of the protocol executes according to one of the bundles
predicted by the semantics.  Moreover, any collection of sessions that
may have occurred takes the following form: its events partition into
bundles that are obtained by instantiating the parameters in bundles
given in the semantics.  Also, if two nodes belong to different
partition elements, there is no $\preceq$ ordering between them,
unless the executions are generated as parts of some higher-level
choreography that might determine a causal ordering.

We call this an {\em abstract bundle semantics} because it builds in
the assumptions of the choreography level: messages do not have
explicit cryptographic operations, and the choreography-level
communication assumptions are satisfied.
%
%
Messages are always delivered exactly once; sender and recipient are
never mismatched; no message is created by adversary operations.  We
must connect this idealized semantics with a more realistic semantics
at the cryptographic level, in which the adversary may be active.

One peculiarity of our message datatype is that we allow ``boxes.''  A
box $\msgbox{\tilde{M}}{\rho\rho'}$ is a message prepared on role
$\rho$ that can be opened only by a principal playing role $\rho'$.
At the choreography level, this property is enforced by a type system.
We use these boxes to make explicit the confidentiality and
authentication requirements of a choreography in the case where some
roles are played by compromised participants.  However, in this
article, we focus on the simplest case, in which no participants are
compromised.  That is, we will assume here, that any participant who
is sent a box, will behave only as predicted by the choreography.

Our semantics at the \emph{cryptographic level} is a standard strand
space treatment, except for one ingredient.  Namely, this semantics
assumes that some kinds of messages are delivered at most once.  These
are session-initiating messages that contain a nonce, or in some
protocols a freshly generated session key.  Implementations now use a
nonce-caching technique in which the nonces of previously executed
sessions are retained in a cache.  A new incoming message contains a
nonce which is compared against the cache; if it is present, then with
overwhelming probability there has been a replay attempt, and the
message is discarded.  Otherwise, the nonce is recorded and the
session proceeds.  So as not to need to retain nonces forever,
implementations typically combine this with a timestamp, and assume
that uncompromised principals are loosely synchronized.  A message
with too old a timestamp is discarded.  Nonces may be dropped from the
cache when their timestamps have expired.  In this approach, the nonce
and the timestamp must appear digitally signed in the incoming message
to prevent manipulation by the adversary.

We define a cryptographic protocol to properly implement a
choreography if, when abstracting its possible executions in this
at-most-once semantics, we obtain exactly the possible executions of
the abstract bundle semantics for the choreography.  

We explore here a simple example in which the participants are
well-known to each other from the start of the transaction.  However,
the ideas also apply when additional participants may be chosen during
execution, and keys must be distributed as part of the message flow.


\section{Strand Spaces}
\label{sec:strandspaces}

Strand spaces \cite{strandspaces,GuttmanEtAl05} were developed as a
simplest possible model for cryptographic protocol analysis, but are
also applicable to other kinds of distributed systems. In strand
spaces, we consider {\em strands}, behavioural traces for roles
represented as finite linear sequences of transmission and reception
events. The model provides techniques for analysing how various
strands can be combined together in a run of a protocol including some
adversary behaviour. 

Let $A$ be a set of messages.
\begin{definition}[Strand Space]
  A {\em directed term} is a pair denoted by $\pm a$ (for $a$ a
  message $\in A$) where $\pm\in\{-,+\}$ is a direction with $+$
  representing transmission and $-$ reception.  A trace is an element
  of $(\pm A)^*$, the set of infinite sequences of directed terms.

  \NI A {\em strand space} is a set $S$ equipped with a trace mapping
  $\mathsf{tr}:S\rightarrow(\pm A)^*$ and its elements are called {\em
    strands}.
\end{definition}
\NI If $s$ is a strand in some strand space $S$ then its
$i^{\mathrm{th}}$ member denotes the $i^{\mathrm{th}}$ transmission or
reception event in $s$.  Formally, we interpret this as the pair
$s,i$, which we call a \emph{node} on the strand $s$.  

We write $m\Rightarrow n$ when, for some $s$ and $i$, $m=s,i$ and
$n=s,i+1$, i.e.~$n$ is the node immediately following $m$ on the
strand $s$.  We write $\msg n$ for the message sent or received in the
directed term of $n$.  That is, if $n=s,i$, and $s(i)$ is a
transmission $+t$ or reception $-t$ of message $t$, then $\msg n=t$.
We occasionally write $\dmsg n=\pm t$ for the message together with
its direction.  We write $m\rightarrow n$ when for some $t$, $\dmsg
m=+t$ and $\dmsg n=-t$.  Thus, $n$ could receive its message directly
from $m$.  

\smallskip

But how can strands be combined together in order to represent
executions of a protocol? This is precisely captured by the notion of
{\em bundle} for a strand space $S$:
\begin{definition}[Bundle] 
  A finite acyclic directed graph
  $\bnd=(\mathcal{N},\mathcal{E},\preceq_{\bnd})$ is a \emph{bundle}
  for $S$ if
  \begin{enumerate}

  \item $\mathcal{N}$ is a set of strand nodes in $S$ such that if
    $n\in\mathcal{N}$ and $m\Rightarrow n$, then $m\in\mathcal{N}$;
    
  \item $\mathcal{E}=\rightarrow_{\bnd}\cup\Rightarrow_{\bnd}$ where
  \begin{enumerate}
    \item $\Rightarrow_{\bnd}$ is the restriction of $\Rightarrow$ to
    nodes in $\mathcal{N}$;
    \item $\rightarrow_{\bnd}\subseteq(\rightarrow\cap\mathcal{N} \times
    \mathcal{N})$; and
    \item for any reception node $n\in\mathcal{N}$, there is exactly
    one transmission node $m\in\mathcal{N}$ such that
    $m\rightarrow_{\bnd}n$.
  \end{enumerate}

  \end{enumerate}
  $n\preceq_{\bnd} m$ iff there is a path using arrows
  $\rightarrow_{\bnd}\cup\Rightarrow_{\bnd}$ from $n$ to $m$ in
  $\bnd$.
\end{definition}
A \emph{bundle} is a causally well-founded graph -- essentially, a
Lamport diagram -- built from strands and transmission edges.  The
relation $\preceq_{\bnd}$ is a well-founded partial order, meaning
that the \emph{bundle induction} principle holds, that every non-empty
set of nodes of $\bnd$ contains $\preceq_{\bnd}$-minimal members.

The notions of strand and bundle, and the principle of bundle
induction, are the essential ingredients in the strand space model.
Choices -- such as 
what operations the adversary strands offer, or what additional
closure properties bundles may satisfy -- can vary to model different
problems concerning cryptographic protocols or distributed
communication more generally.

\smallskip

\NI {\bf Example.}  We briefly introduce an example in order to
clarify the concepts introduced above. Let $S$ be composed by the
following strands:
  \begin{center}
    (1) $n_1\Rightarrow n_2$\quad\qquad (2) $n_3\Rightarrow
    n_4$\quad\qquad (3) $n_5\Rightarrow n_6$\quad\qquad (4)
    $n_7\Rightarrow n_8\Rightarrow n_9\Rightarrow n_{10}$\quad\qquad
    (5) $n_{11}\Rightarrow n_{12}$
  \end{center}
  where
  {\small
  \begin{center}
    \begin{tabular}{llll}
      $\mathsf{dmsg}(n_1)=+\text{``Hello''}$      &
      $\mathsf{dmsg}(n_2)=-\text{``Bye''}$    \\
      $\mathsf{dmsg}(n_3)=+\text{``Good\ luck''}$ &
      $\mathsf{dmsg}(n_4)=-\text{``Thanks''}$ \\
      $\mathsf{dmsg}(n_5)=-\text{``Good\ luck''}$ &
      $\mathsf{dmsg}(n_6)=+\text{``Thanks''}$ \\
      $\mathsf{dmsg}(n_7)=-\text{``Hello''}$      &
      $\mathsf{dmsg}(n_8)=-\text{``Good\ luck''}$ &
      $\mathsf{dmsg}(n_9)=+\text{``Thanks''}$ &
      $\mathsf{dmsg}(n_{10})=+\text{``Bye''}$ \\
      $\mathsf{dmsg}(n_{11})=-\text{``Thanks''}$ &
      $\mathsf{dmsg}(n_{11})=+\text{``Bye''}$
    \end{tabular}
  \end{center}
}
  Below, we report two possible executions in the strand space $S$
  (for clarity, we label $\rightarrow$ with the corresponding message):\\[1mm]
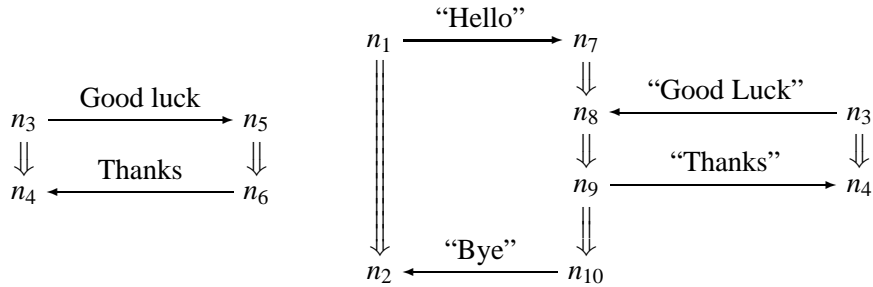
  \begin{center}
    \begin{tabular}{l}
      \begin{diagram}
        n_3       & \rTo^{\quad \text{Good luck}\quad} & n_5\\
        \dImplies &                                    & \dImplies\\\\
        n_4       & \lTo^{\quad\text{Thanks}\quad}  & n_6
      \end{diagram}    
    \end{tabular}
    \qquad
    \begin{tabular}{l}
      \begin{diagram}
        n_1       & \rTo^{\quad\text{``Hello''}\quad}   & n_7\\
        \dImplies &                                   & \dImplies\\\\
                  &                                   & n_8      & \lTo^{\quad\text{``Good Luck''}\quad} & n_3\\
                  &                                   & \dImplies                  &                    & \dImplies\\\\
                  &                                   & n_9      & \rTo^{\quad\text{``Thanks''}\quad} & n_4\\
                  &                                   & \dImplies  \\\\\\
        n_2       & \lTo^{\quad\text{``Bye''}\quad} & n_{10}
      \end{diagram}    
    \end{tabular}
  \end{center}
  Note that strand (5) could interfere allowing for the following
  bundle:
  \begin{center}
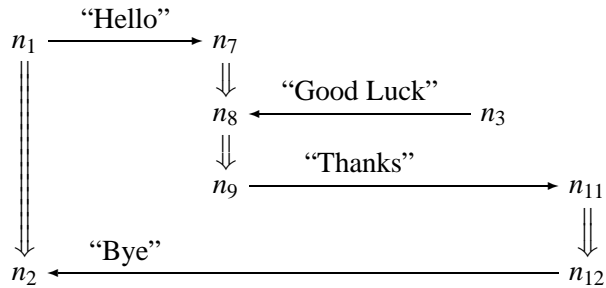
    \begin{tabular}{l}
      \begin{diagram}
        n_1       & \rTo^{\quad\text{``Hello''}\quad}   & n_7\\
        \dImplies &                                   & \dImplies\\\\
                  &                                   & n_8      & \lTo^{\quad\text{``Good Luck''}\quad} & n_3\qquad\\
                  &                                   & \dImplies\\\\
                  &                                   & n_9      & \rTo^{\quad\text{``Thanks''}\quad} &    &   &n_{11}\\
                  &                                   &          &                                     &    &   &\dImplies\\\\\\
        n_2       & \lTo^{\quad\text{``Bye''}\quad} &          &                                     &    &   &n_{12}
      \end{diagram}    
    \end{tabular}
  \end{center}


\section{An Execution Model for Choreography}
\label{sec:choreography}

\subsection{The Calculus}

\paragraph{Syntax.} Let $\rho$ range over the set of roles $\mathcal
R$. The syntax of our choreography mini-language (based on the Global
Calculus \cite{carbone.honda.yoshida:esop07}) is given by the
following grammar:
\begin{align*}
  C::=&\phantom{{}\mid\quad{}}\Sigma_i\,\interact{\rho_1}{\rho_2}{op_i}{\tilde
    M_i}\pfx C_i\mid\quad \INACT\qquad&\qquad M::= &
  \phantom{{}\mid\quad{}}v\mid \msgbox{\tilde
    M}{\rho_1\rho_2} 
\end{align*}
\NI Above, the term $\Sigma_i\,\interact{\rho_1}{\rho_2}{op_i}{\tilde
  M_i}\pfx C_i$ describes an interaction where a branch with label
${\sf op}_i$ is non-deterministically selected and a message $\tilde
M_i$ is sent from role $\rho_1$ to role $\rho_2$. Each two roles in a
choreography share a private channel hence it would be redundant to
have them explicit in the syntax \cite{BCDDDY:concur2008}.

Term $\INACT$ denotes the inactive system. A message $M$ can either be
a value $v$ or a box $\msgbox{\tilde M}{\rho_1\rho_2}$. The latter
denotes a tuple of messages $M_i$ from $\rho_1$ that can only be
opened by $\rho_2$.

\paragraph{Syntactic Assumption.}  The \emph{sender} of a choreography
of the form $\Sigma_i\,\interact{\rho_1}{\rho_2}{op_i}{\tilde M_i}\pfx
C_i$ is $\rho_1$.  We assume, for every choreography $C$:
\begin{itemize}

  \item all $\op{op}$'s are distinct.

  \item in any path in a choreography syntax tree, a box
  $\msgbox{\tilde M}{\rho_1\rho_2}$ has to occur first in an
  interaction whose sender is $\rho_1$ and can only be opened by
  $\rho_2$ in later interaction;

  \item if $C=\Sigma_i\,\interact{\rho_1}{\rho_2}{op_i}{\tilde
    M_i}\pfx C_i$ then either $C_i=\INACT$ or the sender of $C_i$ is
  $\rho_2$ for all $C_i$;

\end{itemize}
The last assumption above requires that the receiving role in an
interaction is always the transmitting role in the subsequent
interaction. All the assumptions above can be statically
checked~\cite{CG09b}.


\paragraph{LTS Semantics.} Our mini-language can be equipped with a
standard trace semantics with configurations $C\lts{\mu} C'$ where
$\mu$ contains the parameters of the interaction performed i.e. it
ranges over the set $\mathcal R\times\mathcal R\times\mathcal
{O}\times \mathcal M$ where $\mathcal O$ is the set of operators
$\op{op}$ and $\mathcal M$ the set of messages.  The following rule
formally defines the relation $\lts{\mu}$ which is taken up to
commutativity and associativity of $+$:
\begin{align*}
  \Did{C-Com}\
  &\
  \Rule
  {}
  {
    \Sigma_i\,\interact{\rho_1}{\rho_2}{op_i}{\tilde M_i}\pfx C_i
    \quad\lts{(\rho_1,\rho_2,\op{op_i},\tilde M)}\quad
    C_i
  }
\end{align*}

\paragraph{Buyer-Seller Example.}
We report a variant of the Buyer-Seller financial protocol
\cite{carbone.honda.yoshida:esop07}. A buyer (or client) $\mathsf{C}$
asks a seller $\mathsf{S}$ for a quote about a product
$\mathsf{prod}$. If the quote is accepted, $\mathsf{C}$ will send its
credit card $\mathsf{card}$ to $\mathsf{S}$ who will forward it to a
bank $\mathsf{B}$. The
bank 
will check if the payment can be done and, if so, reply with a receipt
$\mathsf{receipt}$ which $\mathsf S$ will forward to $\mathsf{C}$.  In
our syntax:
\begin{align*}
  1.\quad &
  \interact{\mathsf{C}}{\mathsf{S}}{\texttt{req}}{\mathsf{prod}}\pfx\
  \interact{\mathsf{S}}{\mathsf{C}}{\texttt{reply}}{\mathsf{quote}}\pfx\\
  2.\quad & (\quad\interact{\mathsf{C}}{\mathsf{S}}{\texttt{ok}}
  {\msgbox{\mathsf{card}}{\mathsf{C}\mathsf{B}}}\pfx\
  \interact{\mathsf{S}}{\mathsf{B}}{\texttt{pay}}
  {\msgbox{\mathsf{card}}{\mathsf{C}\mathsf{B}}}\pfx
  (\quad\interact{\mathsf{B}}{\mathsf{S}}{\texttt{okcf}}
  {\msgbox{\mathsf{receipt}}{\mathsf{B}\mathsf{C}}}\pfx\\
  3.\quad & 
  \phantom{
    (\quad\interact{\mathsf{C}}{\mathsf{S}}{\texttt{ok}}
    {\msgbox{\mathsf{card}}{\mathsf{C}\mathsf{B}}}\pfx\
    \interact{\mathsf{S}}{\mathsf{B}}{\texttt{pay}}
    {\msgbox{\mathsf{card}}{\mathsf{C}\mathsf{B}}}\pfx
    (\quad
  }
  \interact{\mathsf{S}}{\mathsf{C}}{\texttt{rcpt}}
  {\msgbox{\mathsf{receipt}}{\mathsf{B}\mathsf{C}}}\\
  4.\quad & 
  \phantom{
    (\quad\interact{\mathsf{C}}{\mathsf{S}}{\texttt{ok}}
    {\msgbox{\mathsf{card}}{\mathsf{C}\mathsf{B}}}\pfx\
    \interact{\mathsf{S}}{\mathsf{B}}{\texttt{pay}}
    {\msgbox{\mathsf{card}}{\mathsf{C}\mathsf{B}}}\pfx
    (\quad
  }
  \quad\qquad\qquad+\\
  5.\quad & 
  \phantom{
    (\quad\interact{\mathsf{C}}{\mathsf{S}}{\texttt{ok}}
    {\msgbox{\mathsf{card}}{\mathsf{C}\mathsf{B}}}\pfx\
    \interact{\mathsf{S}}{\mathsf{B}}{\texttt{pay}}
    {\msgbox{\mathsf{card}}{\mathsf{C}\mathsf{B}}}\pfx
    (\quad
  }
  \interact{\mathsf{B}}{\mathsf{S}}{\texttt{nopaycf}}{
  }\pfx\\
  6.\quad & 
  \phantom{
    (\quad\interact{\mathsf{C}}{\mathsf{S}}{\texttt{ok}}
    {\msgbox{\mathsf{card}}{\mathsf{C}\mathsf{B}}}\pfx\
    \interact{\mathsf{S}}{\mathsf{B}}{\texttt{pay}}
    {\msgbox{\mathsf{card}}{\mathsf{C}\mathsf{B}}}\pfx
    (\quad
  }
  \interact{\mathsf{S}}{\mathsf{C}}{\texttt{nopay}}{
  }\quad)\\
  7.\quad & \qquad\qquad\qquad\qquad+\\
  8.\quad & \phantom{(\quad}
  \interact{\mathsf{C}}{\mathsf{S}}{\texttt{refuse}}{\mathsf{reason}})
\end{align*} 
\NI Line 1. denotes the quote request and reply. Lines 2. and 8. are
computational branches corresponding to acceptance and rejection of
the quote respectively. If the quote is accepted, $\mathsf{C}$ will
send its credit card in the box
$\msgbox{\mathsf{card}}{\mathsf{C}\mathsf{B}}$ meaning that
$\mathsf{S}$ cannot see it. The card number is then forwarded to
$\mathsf{B}$ who can open the box (line 2.). If the transaction can be
finalised a receipt is forwarded to $\mathsf{C}$.  Otherwise, a
\texttt{nopay} notification will be sent.  $\mathsf{B}$ boxes the
receipt so that it cannot be seen or changed by $\mathsf{S}$.

\subsection{Abstract Bundle Semantics (ABS).}
We introduce an alternative semantics for choreography based on
bundles defined as judgements of the form:
\[\semproves{} C{ \{(\mathcal B_1,\who_1),\ldots,(\mathcal
  B_i,\who_i)\}}\] where $(\mathcal B,\who{})$ is a {\em bundle
  environment}. Given a role $\rho$, $\who(\rho)$ denotes the strand
in the bundle $\mathcal B$ associated to the behaviour of $\rho$.  The
abstract bundle semantics $\sembrack{C} = { \{(\mathcal
  B_1,\who_1),\ldots,(\mathcal B_i,\who_i)\}}$ if and only if
$\semproves{} C{ \{(\mathcal B_1,\who_1),\ldots,(\mathcal
  B_i,\who_i)\}}$.  The relation $\models$ is the minimum relation
satisfying the following:
\begin{displaymath}
  \begin{array}{rl}
    \Did{ABS-Com}\ 
    &
    \Rule
    {
      \begin{array}{l}
        \forall i\pfx\semproves{}{C_i}{\{(\mathcal B_{i1},\who_{i1}),\ldots,(\mathcal B_{ij_i},\who_{ij_i})\}}
      \end{array}
    }
    {
      \semproves{}
      {\Sigma_i\,\interact{\rho_1}{\rho_2}{op_i}{\tilde M_i}\pfx C_i}
      {
        \left(
          \begin{array}{c}
            \bigcup_i \{(\mathcal B_{ij_i},\who_{ij_i})\}_{j_i}[\rho_1,\rho_2,\op{op}_i(\tilde M_i)]\\
          \end{array}
        \right)
      }
    }
  \end{array}
\end{displaymath}
\begin{displaymath}
  \begin{array}{rl}
  \Did{ABS-Zero}\ 
  &
  \Rule
  {
    \op e\text{ fresh}
  }
  {
    \semproves{\emptyset}{\INACT}{(\{\op e^\rho\}_\rho,\lambda \rho\pfx\op e^\rho)}
  }
  \end{array}
\end{displaymath}

\NI The abstract bundle semantics provides a set of bundles which
represents all executions of the protocol described by the
choreography.  In \Did{ABS-Com}, $(\mathcal
B_{ij_i},\who_{ij_i})[\rho_1,\rho_2,\op{op}_i(\tilde M_i)]$ denotes a
new bundle obtained from $\mathcal B_{ij_i}$ where the two strands
$\who_{ij_i}(\rho_1)$ and $\who_{ij_i}(\rho_2)$ are prefixed with the
events $+\op{op}_i(\tilde M_i)$ and $-\op{op}_i(\tilde M_i)$
respectively. The function $\who_{ij_i}$ is updated
accordingly. Formally, 
\begin{equation*}
  (\mathcal B,\who)[\mu]\ = \
  (\quad
  (
  \mathcal N\cup\{n_i\}_i,
  \mathcal E\cup\{n_i\Rightarrow \who(\rho_i)\}_{i}\cup\{n_1\rightarrow n_2\},
  \preceq'
  ),\quad
  \who[\rho_i\mapsto n_i\Rightarrow \who(\rho_i)]_i
  \quad)
\end{equation*}
where $\prec'$ is the update of $\prec_{\mathcal B}$ according to the
new elements added to the bundle and $\mathcal B=(\mathcal N,\mathcal
E,\preceq_{\mathcal B})$. The operation above is applied to all those
bundles obtained from the semantics of each branch and the result will
be their union. In \Did{ABS-Zero}, we augment the set $A$ with fresh
events $\{\op e^\rho\}\in E$ in order to distinguish each strand.

\begin{figure}
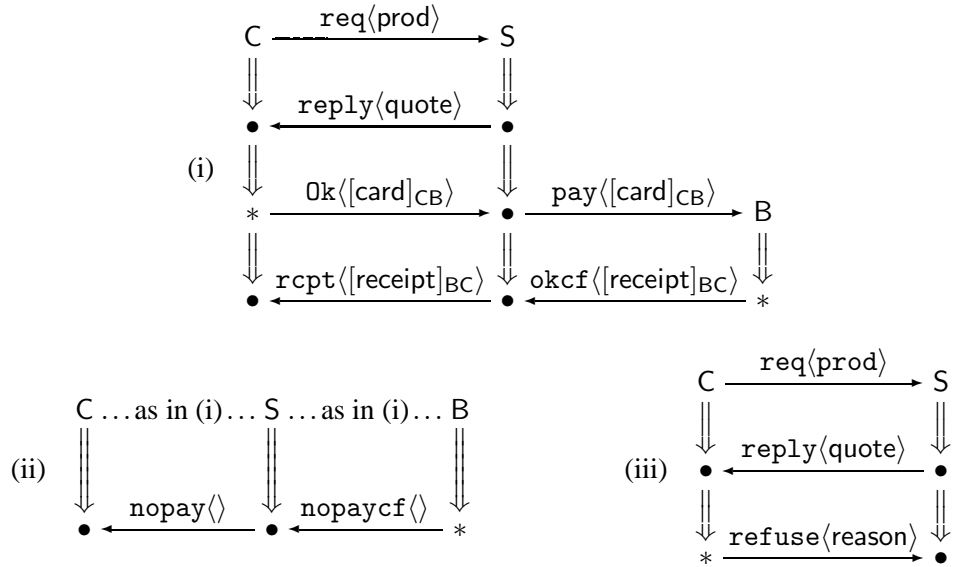

  \centering
    \begin{tabular}{lll}
      (i)\quad
      \begin{diagram}
        \mathsf{C}     & \rTo^{\texttt{req}\langle\mathsf{prod}\rangle}        & \mathsf{S} \\\\\\
        \dImplies      &        &  \dImplies            &                 &   \\
        \bullet        & \lTo^{\texttt{reply}\langle\mathsf{quote}\rangle}          & \bullet        \\\\\\
        \dImplies      &        &  \dImplies            &                 &   \\
        \ast &
        \rTo^{\texttt{Ok}\langle\msgbox{\mathsf{card}}{\mathsf{C}\mathsf{B}}\rangle}&
        \bullet &
        \rTo^{\texttt{pay}\langle\msgbox{\mathsf{card}}{\mathsf{C}\mathsf{B}}\rangle}
        & \mathsf{B} \\\\\\
        \dImplies&&\dImplies&&\dImplies\\
        \bullet & 
        \lTo^{\texttt{rcpt}\langle\msgbox{\mathsf{receipt}}{\mathsf{B}\mathsf{C}}\rangle}
        &\bullet&
        \lTo^{\texttt{okcf}\langle\msgbox{\mathsf{receipt}}{\mathsf{B}\mathsf{C}}\rangle}
        &\ast
      \end{diagram}    
    \end{tabular}
    \\[5mm]
    \begin{tabular}{lll}
      (ii)\quad 
      \begin{diagram}
        \mathsf{C} &
        \ \ldots\text{as in (i)}\ldots\ &
        \mathsf{S} &
        \ \ldots\text{as in (i)}\ldots\ &
        \mathsf{B}
        \\\\\\
        \dImplies&&\dImplies&&\dImplies\\\\\\
        \bullet&\lTo^{\texttt{nopay}\langle\rangle}
        &\bullet&\lTo^{\texttt{nopaycf}\langle\rangle}
        &\ast
      \end{diagram}    
      \quad\qquad&\quad\qquad
      (iii)\quad
      \begin{diagram}
        \mathsf{C} & \rTo^{\texttt{req}\langle\texttt{prod}\rangle}        & \mathsf{S} \\\\\\
        \dImplies      &        &  \dImplies            &                 &   \\
        \bullet        & \lTo^{\texttt{reply}\langle\mathsf{quote}\rangle}          & \bullet        \\\\\\
        \dImplies      &        &  \dImplies            &                 &   \\
        \ast & \rTo^{\texttt{refuse}\langle\mathsf{reason}\rangle} & \bullet
      \end{diagram}    
    \end{tabular}
    \caption{Bundles for the Buyer-Seller protocol}
    \label{fig:exabs}
 \end{figure}

\paragraph{ABS Example.} The ABS for the Buyer-Seller protocol has
three bundles corresponding to its possible executions, namely: (i)
$\mathsf{C}$ accepts the quote and $\mathsf{B}$ successfully finalises
the transaction sending back a receipt; (ii) $\mathsf{C}$ accepts the
quote but $\mathsf{B}$ does not accept the payment; and (iii)
$\mathsf{Buyer}$ does not accept the quote with reason
$\mathsf{reason}$ and the protocol terminates. The three corresponding
bundles are reported in Fig.~\ref{fig:exabs}.
 The nodes marked with $\ast$ are those points where there is a
 possibility of branching i.e.  bundle (ii) is identical to (i) up to
 its $\ast$ while (iii) is identical to (i) and (ii) up to its
 $\ast$. Note that (iii) only involves roles $\mathsf C$ and $\mathsf
 S$.

\smallskip

\smallskip

\NI In the sequel, let $(\mathcal B,\who)\backslash[\mu]$ be defined
as follows: 
\begin{equation*}
  (\mathcal B,\who)\backslash[\mu]\ = \
  \left\{
  \begin{array}{ll}
    \mathcal B'       & \text{ if}\quad\mathcal B = (\mathcal B',\who)[\mu]\\
    \text{undefined}  & \text{ otherwise}
  \end{array}
  \right.
\end{equation*}
Intuitively, the operation above is inverse to $(\mathcal
B,\who)[\mu]$ i.e. removes the first communication from a bundle (if
equal to $\mu$, undefined otherwise). We can then conclude this
section with a result that relates the LTS semantics to the bundle
semantics. 
\begin{theorem}\label{theorem}
  Let $C$ be a choreography. Then, 
  \begin{enumerate}

  \item if $C\lts{\mu}C'$ then there exists a bundle $\mathcal B$ in
    $\sembrack{C}$ such that
    $\sembrack{C'}=\sembrack{C}\backslash(\{\mathcal B\}\cup
    L)\cup\{\mathcal B\backslash[\mu]\}$ for $L = \{\mathcal B'\pp
    \mathcal B\in\sembrack{C}\land\mathcal B\backslash[\mu]\text{ is
      undefined}\}$;

  \item if $\mathcal B\backslash[\mu]$ is defined and $\mathcal
    B\in\sembrack{C}$ then there exists $C'$ such that $C\lts\mu C'$.

  \end{enumerate}
\end{theorem} 

\section{An execution model for Cryptoprotocols}
\label{sec:cryptoprotocols}

Cryptographic protocols are modelled by strand spaces where the set of
messages $a$ is more general. Formally, crypto-level messages, denoted
by the syntactic category $t$ have the following syntax:
\begin{align*}
  t::= & \phantom{{}\mid\quad{}}\tilde v\quad\mid\quad \enc{\tilde
    t}{K}
\end{align*}
Above, the value $v$ ranges over the disjoint union of infinite sets
of nonces (denoted by $N$), atomic keys (denoted by $K$) and other
basic values.  We will write a sequence of messages in the form
$v_1\cons\ldots\cons v_k$. A node of a protocol $\Pi$ is
\emph{regular} if it lies on a strand of $\Pi$, not on an adversary
strand.

\begin{definition}[Deliver-once] Suppose that $S$ is a set of
  messages, and $\bnd$ is a bundle.  $\bnd$ \emph{delivers messages
    in} $S$ \emph{only once} if  there exists an injective
  function $f\colon R\rightarrow T$, where
  \begin{itemize}
  \item $R$ is the set of regular nodes $n$ in $\bnd$ such that a
    member of $S$ is received on $n$, and
  \item $T$ is the set of regular nodes $n$ in $\bnd$ such that a
    member of $S$ is transmitted on $n$.
  \end{itemize} 
%
  When $\{S_i\}_{i\in I}$ is a family of sets indexed by $i\in I$, we
  say that $\bnd$ is \emph{deliver-once} for $\{S_i\}_{i\in I}$ when
  $\bnd$ delivers messages in each $S_i$ only once.
\end{definition} 
We typically apply this definition when $I$ is a set of values that
will be generated freshly, and $S_i$ is a set of messages of
particular forms containing one such value $i$ ($K_{j,k}$ in the
example below).

\begin{figure}
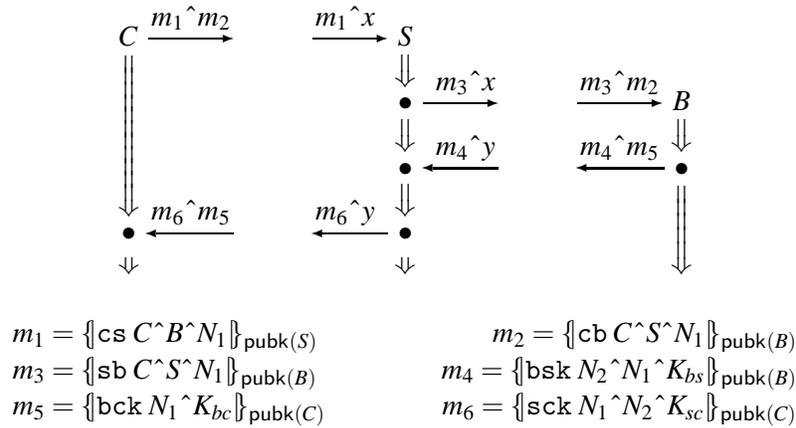

  \centering
  \begin{diagram}[h=3mm,w=8mm]
        C & \rTo{m_1\cons m_2} & \qquad & \rTo{m_1\cons x} & S &&&& \\
        \dStrNext &  & &  & \dStrNext & & & & \\
        &&&& \bullet & \rTo{m_3\cons x}       & \qquad &\rTo{m_3\cons m_2} & B\\
        &&&& \dStrNext & & & & \dStrNext\\
        &&&& \bullet & \lTo{m_4\cons y}&\qquad&\lTo{m_4\cons
          m_5}&\bullet \\
        &&&& \dStrNext &&&& \\ 
        \bullet& \lTo{m_6\cons m_5} &\qquad&\lTo{m_6\cons y} & \bullet
        &&&& \\ 
        \dStrNext &&&& \dStrNext &&&& \dStrNext \\
        \null &&&& \null &&&& \null 
  \end{diagram}
  \begin{tabular}[c]{l@{\qquad\qquad}r}
    $ m_1 = \enc{\texttt{cs}\; C\cons B\cons N_1}{\pk{S}}$ &
    $ m_2 = \enc{\texttt{cb}\; C\cons S\cons N_1}{\pk{B}}$  \\ 
    $ m_3 = \enc{\texttt{sb}\; C\cons S\cons N_1}{\pk{B}}$ &
    $ m_4 = \enc{\texttt{bsk}\; N_2\cons N_1\cons K_{bs}}{\pk{B}}$  \\ 
    $ m_5 = \enc{\texttt{bck}\; N_1\cons K_{bc}}{\pk{C}}$ & 
    $ m_6 = \enc{\texttt{sck}\; N_1\cons N_2\cons K_{sc}}{\pk{C}}$
  \end{tabular}
  \caption{Key exchange phase}
  \label{fig:key:exchange}
\end{figure}

\paragraph{Cryptoprotocol Example.} The Buyer-Seller cryptoprotocol
implements the choreography example of Section~\ref{sec:choreography}.
It provides parametric strands that define the behaviors of the
principals as they send and receive encrypted messages to provide
security services for the behaviors in the choreography.  The central
idea is that the first few messages use public encryption keys and
nonces to establish symmetric keys.  The remaining messages then use
the keys in a straightforward way.  To establish a key between $A$ and
$B$, $A$ sends a message containing a nonce, encrypted with $B$'s
public key.  $B$ returns a message encrypted with $A$'s public key.
It contains $A$'s nonce as well as a fresh symmetric key to be used
for this session.  We use different syntactic tags in each encrypted
unit which correspond to the $\op{op}$'s in the choreography (denoted
by the typewriter font \texttt{op}).  At this level, the tags ensure
that no unit can be confused with any other (this is the reason why
the $\op{op}$'s are all distinct at choreography level).  
The key exchange phase takes the form shown in
Fig.~\ref{fig:key:exchange}.
Each participant leaves the key exchange phase knowing that $N_1,N_2$
are shared among $C,S,B$, and that two symmetric keys are to be used
for encryption in the next phase.  For instance, $C$ knows to use
$K_{sc}$ to communicate with the seller in the ensuing exchange, and
to use $K_{bc}$ to communicate with the bank.  

\begin{figure}[th]
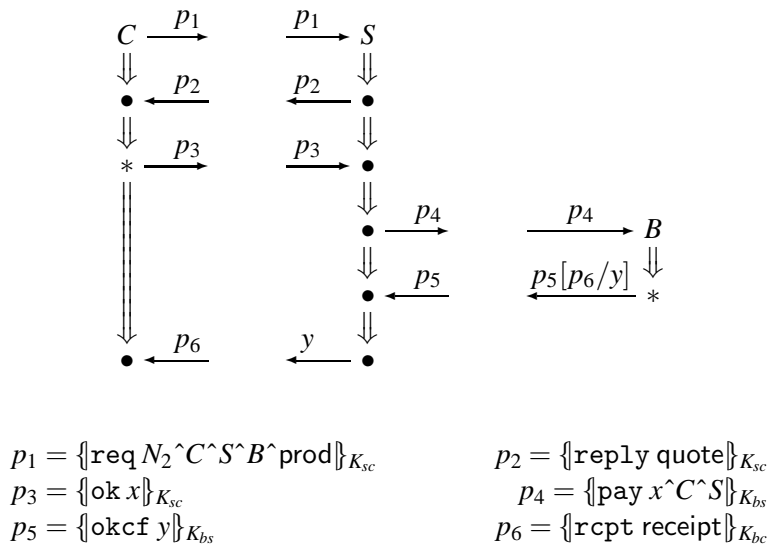

  \centering
    \begin{diagram}[h=3mm,w=8mm]
      C & \rTo{p_1} & \qquad & \rTo{p_1} & S &&&& \\
      \dStrNext &  & &  & \dStrNext & & & & \\
      \bullet & \lTo{p_2} & \qquad & \lTo{p_2} & \bullet &&&& \\
      \dStrNext &  & &  & \dStrNext & & & & \\
      \ast & \rTo{p_3} & \qquad & \rTo{p_3} & \bullet &&&& \\
      \dStrNext &  & &  & \dStrNext & & & & \\
      &&&& \bullet & \rTo{p_4}       & \qquad &\rTo{p_4} & B\\
      &&&& \dStrNext & & & & \dStrNext\\
      &&&& \bullet & \lTo{p_5}&\qquad&\lTo{p_5[p_6/y]} & \ast \\
      &&&& \dStrNext &&&& \\
      \bullet& \lTo{p_6} &\qquad&\lTo{y} & \bullet
      &&&& 
  \end{diagram}
  \ \\
  \ \\
  \ \\
  \begin{tabular}[c]{l@{\qquad\qquad}r}
    $ p_1 = \enc{\texttt{req}\; N_2 \cons C\cons S\cons B\cons \mathsf{prod}}{K_{sc}}$ &
    $ p_2 = \enc{\texttt{reply}\; \mathsf{quote}}{K_{sc}}$  \\ 
    $ p_3 = \enc{\texttt{ok}\; x}{K_{sc}}$ &
    $ p_4 = \enc{\texttt{pay}\; x \cons C\cons S}{K_{bs}}$  \\ 
    $ p_5 = \enc{\texttt{okcf}\; y}{K_{bs}}$ & 
    $ p_6 = \enc{\texttt{rcpt}\; \mathsf{receipt}}{K_{bc}}$
  \end{tabular}
  \caption{Payload exchange phase}
  \label{fig:payloads}
\end{figure}

In the ensuing stage, the participants use these keys to transfer the
payloads amongst themselves.  Their exchange---in the successful case,
in which the transaction completes---takes the form shown in
Fig.~\ref{fig:payloads}.
However, $\mathsf{C}$ and $B$ each have an opportunity to prevent the
exchange from completing, at the nodes marked $\ast$.  If $C$
transmits $\enc{\texttt{refuse}}{K_{sc}}$ instead of $p_3$, then $S$
must terminate the exchange before contacting $B$.  If $B$ transmits
$\enc{\texttt{nopaycf}\; \enc{\texttt{nopay}}{K_{bc}}}{K_{bs}}$ instead
of $p_5[p_6/y]$, then $S$ and $C$ must terminate the transaction.

Let us assume that the participants of a run use their private
decryption keys only in accordance with this protocol, and that the
nonces $N_1,N_2$ and keys $K_{bc},K_{bs},K_{sc}$ are in fact freshly
chosen and unguessable.  On this assumption, there are essentially
only three possible executions, if we consider only those of minimal
size, given that a role completed.  When $C$ completes normally, then
the other participants have completed normally with matching
parameters.  When $S$ completes with a client refusal, then $C$ has
refused and $B$ has had a matching key exchange phase but no more.
When $C$ completes with a $\texttt{nopay}$ message, then $B$ has
refused to pay, and $S$ has been informed of this.  This analysis
indicates that the protocol appears to achieve its goals.  Indeed, we
have confirmed this with the tool \textsc{cpsa}, a Cryptographic
Protocol Shapes Analyzer~\cite{DoghmiGuttmanThayer07}, which
enumerates the minimal, essentially different executions of the
protocol.  We can then check the assertions we have just made by
inspecting those executions.

\section{Abstraction and Correctness} A partial function $\alpha$ over
messages is an \emph{abstraction map} if (1) $\alpha(t)$ (if defined)
contains no cryptographic operators, nonces nor keys, and (2) the
parameters in $\alpha(t)$ (if defined) always appear in $t$.

For instance, $\alpha$ could map $\enc{\texttt{req}\; N_2 \cons C\cons
  S\cons B\cons \mathsf{prod}}{K_{sc}}$ to
${\texttt{req}\langle\mathsf{prod}\rangle}$ in our Buyer-Seller
example.  The result has no cryptography and no nonces, and the tags
$\texttt{req}$ and $\mathsf{prod}$ appear in the argument.  

We say that an abstract strand $s$ is an \emph{image} of a
cryptographic strand $s_c$ if, ignoring transmissions or receptions on
$s_c$, for which $\alpha$ is undefined, for each transmission or
reception node $n$ on $s$, its message $\term(n)$ is
$\alpha(\term(n_c))$, where $n_c$ is the corresponding transmission or
reception node (resp) on $s_c$.  That is, $\alpha$ yielding the trace
of $s$, when mapped through the trace of $s_c$ restricted to the
domain of $\alpha$.

Suppose that a concrete strand $s_c$ has its first $i$ nodes in a
concrete bundle $\bndC$, but $\alpha$ is undefined for the messages on
these nodes.  We then say that $s_c$ is \emph{abstractly vacuous in}
$\bndC$.  In the opposite case, when some node $n$ of $s_c$ is in
$\bndC$ and $\alpha(\term(n))$ is well-defined, we say that $s_c$ is
\emph{abstractly non-vacuous in} $\bndC$.

An abstract bundle $\bnd$ is an \emph{image} of a cryptographic bundle
$\bndC$ if (1) there is a bijection $\phi$ between the abstractly
non-vacuous regular strands $s_c$ of $\bndC$ and the regular strands
$s$ of $\bnd$; (2) $\phi(s_c)$ is always an image of $s_c$; and (3)
the transmission relation $\rightarrow_{\bnd}$ is formed by connecting
nodes of $\bnd$ such that $m\rightarrow_{\bnd}n$ implies
$m_c\preceq_{\bndC}n_c$, for some concrete nodes of which $m,n$ are
images.  See~\cite{Guttman09a} for a related notion of protocol
transformation, and~\cite{MaffeiEtAl07} for an approach to protocol
verification via abstraction functions.

Suppose that $\bndC$ is a concrete bundle and $\{\bndC_i\}_i$ is a
family of sub-graphs of $\bndC$ that partitions the regular nodes of
$\bndC$.  We say that $\{\bndC_i\}_i$ \emph{separates} $\bndC$
\emph{into components} when each $\bndC_i$ is a bundle on its own.

\begin{definition}[Faithfulness] \label{def:faithful}
Cryptoprotocol $\Pi$ is \emph{faithful to} choreography $C$ if there
is an abstraction function $\alpha$ such that:
\begin{enumerate}
  \item Every $\bnd\in\sembrack{C}$ is an image of some bundle $\bndC$
  of $\Pi$;\label{clause:fth:upward:cover}
  \item If $\bndC$ is a bundle of $\Pi$, then some family
  $\{\bndC_i\}_i$ separates $\bndC$ into components.  Moreover, each
  image $\bnd_i$ of any $\bndC_i$ is an initial sub-bundle of
  $\sigma(\bnd)$, for some $\bnd\in\sembrack{C}$ and some substitution
  $\sigma$.\label{clause:fth:downward:separate}
\end{enumerate}
If $\{S_i\}_{i\in I}$ is a family of sets of messages, then $\Pi$ is
\emph{faithful to} $C$ \emph{assuming the deliver-once property for}
$\{S_i\}_{i\in I}$ if the above holds for bundles of $\Pi$ that are
deliver-once for $\{S_i\}_{i\in I}$.
\end{definition}

\smallskip

\NI {\bf Faithfulness in the Buyer-Seller protocol.} We use the
protocol analysis tool \textsc{cpsa}~\cite{DoghmiGuttmanThayer07} as
part of a proof that the protocol of Fig.~\ref{fig:key:exchange} and
Fig.~\ref{fig:payloads} is faithful to the choreography in
Fig.~\ref{fig:exabs}.  There are three stages:
\begin{enumerate}
  \item \textsc{cpsa} determines the minimal, essentially different
  executions that are possible, given that any one party has had a
  complete run.  

  These are the expected success execution $\skel_s$ and failure
  execution $\skel_f, \skel_{f'}$, modulo the fact that a party never
  knows whether its last message was successfully delivered, if its
  last action is a transmission.  In particular, the active parties
  agree on all parameters to the session.
  \item Based on this \textsc{cpsa} output, inspection shows that
  Def.~\ref{def:faithful}, Clause~\ref{clause:fth:upward:cover} is
  satisfied:  Any run $\bnd\in\sembrack{C}$ is the abstraction of some
  concrete bundle $\bndC$.
  \item Because $\skel_s,\skel_f,\skel_{f'}$ are the only minimal
  forms of execution, every larger execution $\bnd_c$ is a (possibly
  non-disjoint) union of executions of these forms.  That is, there is
  a family of maps $\{H_i\}_i$, where each $H_i$ maps either $\skel_s$
  or $\skel_f$ to some subset of the regular nodes of $\bnd_c$.
  Moreover, each regular node $n\in\bnd_c$ is the image of some node
  in $\skel_s,\skel_f$, or $\skel_{f'}$ under at least one of the
  $H_i$.

  However, each pair of strands agrees on a pair of freshly chosen
  values, where each of them has chosen one of the values.  This
  forces the range of $H_i$ and $H_j$ either to coincide or be
  disjoint.  Hence Clause~\ref{clause:fth:downward:separate} is
  satisfied when we define the family $\{\bndC_i\}_i$ by saying that
  two nodes belong to the same $\bndC_i$ if they are both in the range
  of any one $H_i$.    
\end{enumerate}


\section{Concluding Remarks}
We have introduced two execution models, one for choreography
(assuming no compromised participants) and one for cryptoprotocols
with deliver-once assumptions.  The abstract bundle semantics gives a
set of bundles representing all the possible runs of the protocol
described by a choreography.  We have sketched a form of argument for
proving that a cryptoprotocol is faithful to the ABS of a
choreography.

In \cite{CG09b}, we studied an abstract semantics for the choreography
language presented here where roles can belong to compromised
principals. The ideas of abstraction have yet to be extended to the
compromised case and to a choreography language with infinite
states. 
The work by Bhargavan et al. in \cite{BhargavanEtAl09, CDFBL08} is
closely related to ours: they provide a compiler for generating ML
code that can then be type-checked for verifying its security
property. Their notion of faithfulness is guaranteed for the
well-typed code generated from the source choreography.

In future work, we aim at developing systematic techniques for proving
that certain transformations preserve all of the goals of a protocol,
while achieving additional goals~\cite{Guttman09a}.


%
\label{sect:bib}
\bibliographystyle{plain}
\bibliography{session}


\end{document}